\documentclass[prd,onecolumn,superscriptaddress,showpacs,preprintnumbers,amsmath,amssymb,nofootinbib]{revtex4}
\usepackage{dcolumn}
\usepackage{bm}
\usepackage[latin1]{inputenc}
\usepackage[spanish,english]{babel}
\usepackage{amsfonts}
\usepackage{amssymb}
\usepackage{graphicx}

\newcommand{\be}{\begin{equation}}
\newcommand{\ee}{\end{equation}}
\newcommand{\bea}{\begin{eqnarray}}
\newcommand{\eea}{\end{eqnarray}}

\begin{document}
\title{{\bf Inflation, Quantum Field Renormalization, and CMB Anisotropies }\footnote{Talk given in the 12th Marcel Grossman Meeting (Paris, July 2009). To appear in the proceedings.}}
\author{Iván Agulló}\email{ivan.agullo@uv.es}
 \affiliation{ {\footnotesize Physics Department, University of
Wisconsin-Milwaukee, P.O.Box 413, Milwaukee, WI 53201 USA}}
\author{José Navarro-Salas}\email{jnavarro@ific.uv.es}
\affiliation{ {\footnotesize Departamento de Física Teórica and
IFIC, Centro Mixto Universidad de Valencia-CSIC.
    Facultad de Física, Universidad de Valencia,
        Burjassot-46100, Valencia, Spain. }}

\author{Gonzalo J. Olmo}\email{olmo@iem.cfmac.csic.es }
\affiliation{{\footnotesize Instituto de Estructura de la Materia,
CSIC, Serrano 121, 28006 Madrid, Spain}}\affiliation{ {\footnotesize Physics Department, University of
Wisconsin-Milwaukee, P.O.Box 413, Milwaukee, WI 53201 USA}}

\author{Leonard Parker}\email{leonard@uwm.edu}
\affiliation{ {\footnotesize Physics Department, University of
Wisconsin-Milwaukee, P.O.Box 413, Milwaukee, WI 53201 USA}}

\date{May 14th, 2010}

\begin{abstract}
We point out that if quantum field renormalization is taken into account the predictions of slow-roll inflation for both the scalar and tensorial power spectra change significantly for wavelengths that today are at observable scales.
\\

\end{abstract}


\maketitle

Inflation  provides a quantitative
explanation \cite {books} to account for the origin of small inhomogeneities in the early universe.
The potential-energy density of the inflaton field
is assumed to cause the inflationary  accelerated expansion, and the
amplification of its quantum fluctuations and those of the metric are inevitable consequences in an
expanding universe \cite{parker69}. These fluctuations  acquire classical properties during the inflationary period and  provide the initial conditions
for classical cosmological perturbations after the big-bang.
The detection of the effects of primordial tensorial metric fluctuations (gravitational waves) in future high-precision measurements of the
CMB anisotropies
will  serve as a highly non-trivial test of the inflationary paradigm and to constrain specific models. Therefore,
it is particularly important to scrutinize the predictions of inflation for the tensorial and scalar power spectra. In this respect, it was pointed out in \cite{Parker07} (see also \cite{agulloetal08}) that quantum field renormalization significantly modifies  the amplitude of quantum fluctuations, and hence the corresponding power spectra, in de Sitter inflation. The analysis was further improved in \cite{agulloetal09, agulloetal092} to
 understand how the basic testable predictions of (single-field) slow-roll inflation
 could be affected by  renormalization. In this talk we summarize our approach.

Let us assume that $\varphi(\vec{x},t)$ represents a perturbation obeying a free field wave-equation on the inflationary background $ds^2= -dt^2+a^2(t)d\vec{x}^2$, where $a(t)$ is a quasi-exponential expansion factor. At the quantum level, this field is expanded as
$\varphi(\vec{x},t)= (2\pi)^{-3/2} \int d^3k [\varphi_{k}(t)a_{\vec{k}}e^{i\vec{k}\vec{x}} +  \varphi_{k}^*(t)a^{\dagger}_{\vec{k}}e^{-i\vec{k}\vec{x}}]$,
where the creation and annihilation operators satisfy the canonical commutation relation.
The mode functions $\varphi_{k}(t)$ are required to satisfy the adiabatic condition \cite{parker-toms}.
 The  power spectrum  for this perturbation, $\Delta^2_{\varphi}(k,t)$, is usually defined as \cite{books}
$\langle \hat \varphi_{\vec{k}}(t) \hat \varphi^{\dagger}_{\vec{k}'}( t)\rangle =  \delta^3 (\vec{k} - \vec{k}')\frac{2\pi^2}{k^3}\Delta^2_{\varphi}(k,t)$,
where $\hat \varphi_{\vec{k}}(t)\equiv\varphi_{k}(t)a_{\vec{k}}$.
These modes describe a perturbation field characterized, in momentum space, by a zero mean $\langle \hat \varphi_{\vec{k}}(t)\rangle =0$ and the variance $\Delta^2_{\varphi}(k,t)$. The advantage of working in momentum space resides in the fact that different modes fluctuate independently of each other. This way, the quantum field is regarded as an infinite collection of independent oscillators, each with a different value of $\vec{k}$. In position space the perturbation is also characterized by a zero mean $\langle \varphi(\vec{x},t)\rangle=0$ and a variance $ \langle \varphi^2(\vec{x},t) \rangle = (2\pi)^{-3}\int d^3k d^3k' \langle \hat \varphi_{\vec{k}}(t) \hat \varphi^{\dagger}_{\vec{k}'}(t)\rangle e^{i (\vec{k}-\vec{k'}) \vec{x}}$.
 This variance is formally related to the power spectrum  by
\be \label{tpfunctioncp}\langle \varphi^2(\vec{x},t) \rangle = \int_0^{\infty}\frac{dk}{k} \Delta^2_{\varphi}(k,t) \sim \frac{1}{4\pi^2}\int_0^{\infty}dk(\frac{k}{a^2} + \frac{\dot a^2}{a^2k} + ...)  \ , \ee where the large $k$ behavior of the integrand is shown.
As is well-known in quantum field theory, the above  expectation value quadratic in the field $\varphi$ is ultraviolet divergent (quadratic and logarithmically).
The quadratic divergence corresponds to the usual contribution from vacuum fluctuations in Minkowski space and can be eliminated by standard renormalization in flat spacetime. The logarithmic divergence, however, appears as a consequence of the non trivial expansion. Because the different $k$-modes fluctuate independently of each other, one could be tempted to get rid of this logarithmic ultraviolet divergence by simply eliminating the modes with $k>aH$ and leaving the rest  unaffected. One then obtains $\Delta^2_{\varphi}(k)\approx H^2/4\pi^2$, where $\Delta^2_{\varphi}(k)$ is defined by the quantity  $\Delta^2_{\varphi}(k,t)$ evaluated a few Hubble times  after the ``horizon crossing time'' $t_k$ ($a(t_k)/k=H(t_k)$), since this is the time scale at which the modes behave as classical perturbations. However,  there are fundamental holistic aspects of quantum field theory (QFT) that can not be properly understood in terms of independent $k$-modes. Renormalization is the hallmark of the holistic aspects of QFT \cite{hollands-wald}. Therefore, the logarithmic divergence in (\ref{tpfunctioncp}) should be dealt with by renormalization in curved spacetimes and we propose that in the standard definitions of the spectrum $\Delta^2_{\varphi}(k, t)$, as given in  (\ref{tpfunctioncp}), one should replace the unrenormalized $\langle \varphi^2(\vec{x},t) \rangle$ by the renormalized variance, $\langle \varphi^2(\vec{x},t) \rangle_{ren}$. Writing $\tilde\Delta^2_{\varphi}(k,t)$ for the spectrum defined in this way,  the definition in (\ref{tpfunctioncp}) is replaced by the corresponding renormalized expression
\be \label{varren}\langle \varphi^2(\vec{x},t) \rangle_{ren}   = \int_0^{\infty}\frac{dk}{k} \tilde\Delta^2_{\varphi}(k,t) \ . \ee
Since the power spectrum is defined in momentum space, the natural scheme is renormalization in momentum space, so we define
 \be \label{varrendef}\langle \varphi^2(\vec{x},t)\rangle_{ren}   = \frac{4\pi}{(2\pi)^3}\int_0^{\infty}k^2dk(|\varphi_{k}(t)|^2 - C_{k}(t)) \ , \ee
where $C_{k}(t)$ represents the renormalization counterterms. Adiabatic renormalization \cite{parker-fulling, parker-toms} provides a natural expression for
$C_{k}(t)$. Moreover, the Bunch-Parker renormalization in momentum space \cite{Bunch-Parker, parker-toms} (which turns out to be equivalent, when translated to position space, to the DeWitt-Schwinger proper time renormalization) provides another answer for $C_{k}(t)$.  When these schemes are applied to the field perturbations arising from slow-roll inflation, which should be considered as massless free fields with a second-order adiabatic term encoding the dependence on the inflationary potential, the resulting expressions for  $C_{k}(t)$ coincide, thus defining a unique expression for  the spectrum $\tilde\Delta^2_{\varphi}(k,t)$.
The holistic nature of QFT is then explicitly realized through (\ref{varren}-\ref{varrendef}); although the counterterms are fully determined by the ultraviolet behavior of the modes,  the long wavelength sector, and hence the new $\tilde\Delta^2_{\varphi}(k,t)$, is significantly affected by the  subtractions.
In the slow-roll scenario, when $H$ slowly decreases with time, the effects of renormalization have a non-trivial impact on $\tilde\Delta^2_{\varphi}(k,t)$ when this quantity is evaluated a few $n$ Hubble times after the time of horizon crossing $t_k$ ($n>1, n\epsilon \ll 1$). We obtain for tensorial and scalar spectra\bea \tilde\Delta_t^2(k,n)&\approx & \frac{2}{M_P^2}\left (\frac{H(t_k)}{2\pi} \right )^2 \epsilon(t_k) (2n-3/2) \nonumber \\ \tilde\Delta_{s}^2(k,n)&\approx & \frac{1}{2M_P^2\epsilon(t_k)}\left
(\frac{H(t_k)}{2\pi}\right)^2 (3\epsilon(t_k) - \eta(t_k))(2n-3/2)\ , \label{eq:DR-n}\eea
 where $\epsilon, \eta$ are the standard slow-roll parameters.
Note that  the parameter $n$ enters in the power spectra parameterizing the (unknown) time at which the modes exhibit classical behavior.
However,
the tensor-to-scalar ratio $r$ is not sensitive to the unknown parameter $n$. 
As a consequence of (\ref{eq:DR-n}), the imprint of slow-roll inflation on the CMB anisotropies is significantly altered. Note that, in contrast with the standard (unrenormalized) approach, the tensorial amplitude does not uniquely depend  on the scale  of inflation. Moreover, one can also obtain a non-trivial  change in the consistency condition that relates the tensor-to-scalar ratio $r$ to the spectral indices \cite{agulloetal092}. For instance, an exact scale-invariant tensorial power spectrum, $n_t=0$, is now compatible with a non-zero ratio $r\approx 0.12\pm0.06$, which is forbidden by the standard prediction ($r=-8n_t$). {\it Acknowledgements.} This
work has been partially supported by the  grants FIS2008-06078-C03-02, PHY-0503366 and the ``Jos\'e Castillejo'' program. G.O.
thanks MICINN for a JdC contract.

\end{document}